\documentclass[amsmath,aps,showpacs,a4paper,10pt]{revtex4}

 \usepackage{epsf}
 \usepackage{graphicx}    

\begin{document}

 \newcommand{\be}[1]{\begin{equation}\label{#1}}
 \newcommand{\ee}{\end{equation}}
 \newcommand{\bea}{\begin{eqnarray}}
 \newcommand{\eea}{\end{eqnarray}}
 \def\disp{\displaystyle}

 \def\gsim{ \lower .75ex \hbox{$\sim$} \llap{\raise .27ex \hbox{$>$}} }
 \def\lsim{ \lower .75ex \hbox{$\sim$} \llap{\raise .27ex \hbox{$<$}} }

 \begin{titlepage}

 \begin{flushright}
 arXiv:0808.2240
 \end{flushright}

 \title{\Large \bf Reconstructing the Cosmic Expansion History
 up to Redshift $z=6.29$ with the Calibrated Gamma-Ray Bursts}

 \author{Hao~Wei\,}
 \email[\,email address:\ ]{haowei@bit.edu.cn}
 \affiliation{Department of Physics, Beijing Institute
 of Technology, Beijing 100081, China}

 \author{Shuang~Nan~Zhang}
 \affiliation{Department of Physics and Tsinghua Center for
 Astrophysics,\\ Tsinghua University, Beijing 100084, China\\
 Key Laboratory of Particle Astrophysics, Institute of High
 Energy Physics,\\
 Chinese Academy of Sciences, Beijing 100049, China\\
 Physics Department, University of Alabama in Huntsville,
 Huntsville, AL 35899, USA}

 \begin{abstract}\vspace{1cm}
 \centerline{\bf ABSTRACT}\vspace{2mm}
Recently, Gamma-Ray Bursts (GRBs) were proposed to be a
 complementary cosmological probe to type Ia supernovae (SNIa).
 GRBs have been advocated to be standard candles since several
 empirical GRB luminosity relations were proposed as distance
 indicators. However, there is a so-called circularity problem
 in the direct use of GRBs. Recently, a new idea to calibrate
 GRBs in a completely cosmology independent manner has been
 proposed, and the circularity problem can be solved. In the
 present work, following the method proposed by Liang {\it et al.},
 we calibrate 70 GRBs with the Amati relation using 307 SNIa.
 Then, following the method proposed by Shafieloo {\it et al.},
 we smoothly reconstruct the cosmic expansion history up
 to redshift $z=6.29$ with the calibrated GRBs. We find some
 new features in the reconstructed results.
 \end{abstract}


 \maketitle

 \end{titlepage}

 \renewcommand{\baselinestretch}{1.6}



\section{Introduction}\label{sec1}
The current accelerated expansion of our universe~\cite{r1} has been
 one of the most active fields in modern cosmology since its
 discovery in 1998 from the observations of type Ia supernovae
 (SNIa)~\cite{r2}. Later, the observations of cosmic microwave
 background (CMB) anisotropy~\cite{r3} and large-scale
 structure~\cite{r4} confirmed this discovery. Although today there
 are already many observational methods, SNIa have been proved to be
 one of the most powerful tools to probe this mysterious phenomenon.
 In 1992, the famous Phillips relation was found~\cite{r5}, which
 claimed that for nearby SNIa there exists a clear correlation
 between their intrinsic brightness at maximum light and the
 duration of their light curve. Since then, some empirical
 technics~\cite{r6,r7,r8} have been developed to use the Phillips
 relation to make SNIa into standard candles. See
 e.g.~\cite{r9,r10} for brief historical reviews.

However, SNIa are plagued with extinction from the interstellar
 medium. Hence, the currently maximum redshift of SNIa is only about
 $z\simeq 1.755$. As argued in~\cite{r10}, the observations at
 $z>1.7$ are fairly important to distinguish cosmological models. On
 the other hand, the redshift of the last scattering surface of CMB
 is $z\simeq 1090$. So, the observations at intermediate redshift
 are important. Recently, Gamma-Ray Bursts (GRBs) were proposed
 to be a complementary probe to SNIa. So far, GRBs are the most
 intense explosions observed in our universe. Their high energy
 photons in the gamma-ray band are almost immune to dust extinction.
 Up to now, there are many GRBs observed at $0.1<z\leq 8.1$, whereas
 the maximum redshift of GRBs is expected to be $10$ or even
 larger~\cite{r11}. Strictly speaking, GRBs are not standard
 candles, with radiated energies spanning several orders of
 magnitude. But the use of empirical luminosity correlations to
 standardize them as distance indicators has been proposed by
 several authors. In some sense, this is reminiscent of the
 case of the Phillips relation for SNIa. These empirical
 correlations include the Amati relation~\cite{r12}, those
 derived from it (for examples, the Ghirlanda relation~\cite{r13},
 the Yonetoku relation~\cite{r14}, the Liang-Zhang
 relation~\cite{r15}, the Firmani relation~\cite{r16}), and
 others~\cite{r17,r18,r19,r20}. Therefore, the so-called GRB
 cosmology emerges recently. We refer to
 e.g.~\cite{r21,r22,r23,r10} for comprehensive reviews.

To our knowledge, in~\cite{r24} GRBs were used to constrain
 $\Omega_m$ and dark energy for the first time. Similar works also
 include~\cite{r25} for examples. However, there
 is a so-called circularity problem in the direct use of
 GRBs~\cite{r21}. In this case, to calibrate the empirical GRB
 luminosity relations, one need to assume a particular cosmological
 model with some model parameters {\it a priori}, mainly due to the
 lack of a set of low redshift GRBs at $z<0.1$ which are
 cosmology independent. When one uses these ``calibrated'' GRBs
 (which are actually cosmological model dependent) to constrain
 cosmological models, the circularity problem occurs. To alleviate
 the circularity problem, some statistical methods have been
 proposed, such as the scatter method~\cite{r26}, the luminosity
 distance method~\cite{r26}, and the Bayesian method~\cite{r27}.
 These methods were used extensively in the
 literature~\cite{r15,r28,r29}. However, they still cannot solve
 the circularity problem {\em completely}. Another method trying
 to avoid the circularity problem was proposed in~\cite{r30},
 in which the parameters of the empirical GRB luminosity relation
 and the cosmological model parameters were fitted to the
 observational data simultaneously. However, for {\em any} given
 cosmological model, this method can {\em always} obtain
 some parameters for the cosmological model and the empirical
 GRB luminosity relation. In this sense, {\em any} cosmological
 model is ``viable'' (except for a few obviously absurd models). So,
 it cannot be used to rule out any cosmological model.
 Therefore, it is also an unsatisfactory method to solve the
 circularity problem completely.

To overcome the circularity problem completely, one should calibrate
 GRBs in a cosmology independent manner. Due to the lack of a set
 of low redshift GRBs at $z<0.1$ which are cosmology independent,
 it was proposed to calibrate the empirical GRB luminosity relation
 using a sufficient number of GRBs within a small redshift bin
 centered around any redshift~\cite{r31} (see also e.g.~\cite{r21}).
 However, this method might be unrealistic, since the current sample
 of observed GRBs is not large enough. Recently, a new idea to
 calibrate GRBs in a completely cosmology independent manner has
 been proposed in~\cite{r32,r33} independently. Similar to the case
 of calibrating SNIa as secondary standard candles by using Cepheid
 variables which are primary standard candles, we can also calibrate
 GRBs as standard candles with a large amount of SNIa. This idea of
 the distance ladder was also briefly mentioned in~\cite{r34}.
 However, in fact, \cite{r34} used the $\Lambda$CDM model
 rather than SNIa data themselves to calibrate GRBs and hence
 the calibration is still cosmology dependent actually.
 In~\cite{r32}, the GRB luminosity relation was calibrated with
 the empirical formula of the luminosity distance of SNIa,
 while the physical meaning of this empirical formula is still
 unclear and its reliability should be tested carefully. Instead,
 in~\cite{r33} the GRB luminosity relation was calibrated with
 the interpolated distance moduli from the Hubble diagram of
 SNIa themselves. We refer to the original papers~\cite{r32,r33} for
 details. Using these cosmology independently calibrated GRBs,
 we can constrain the parameters of cosmological models without
 circularity problem. In fact, the constraints on $\Lambda$CDM model
 and $w_0-w_a$ parameterized dark energy model have been considered
 in~\cite{r32,r33,r35}.

In the present work, we will calibrate GRBs to reconstruct the
 cosmic expansion history up to redshift $z=6.29$. The calibration
 method of GRB relation adopted here is the one proposed
 in~\cite{r33}. However, we will use the datasets of SNIa and GRBs
 which are larger than the ones used in~\cite{r33}. So, more
 calibrated high redshift GRBs can be available with
 smaller uncertainties. In Sec.~\ref{sec2}, we use 38 GRBs
 at $z<1.4$ to calibrate the Amati relation while the
 distance moduli of these 38 low redshift GRBs are interpolated
 from 307 SNIa data~\cite{r36,r37}. Then, the distance moduli of
 32 GRBs whose $z>1.4$ can be derived from the calibrated Amati
 relation. In Sec.~\ref{sec3}, following the method proposed
 in~\cite{r38,r39}, we smooth 307 SNIa data and 32 calibrated
 GRBs data whose $z>1.4$ to reconstruct the cosmic expansion
 history up to $z=6.29$. Conclusion and discussions are briefly
 given in Sec.~\ref{sec4}.


 \begin{center}
 \begin{figure}[htbp]
 \centering
 \includegraphics[width=0.99\textwidth]{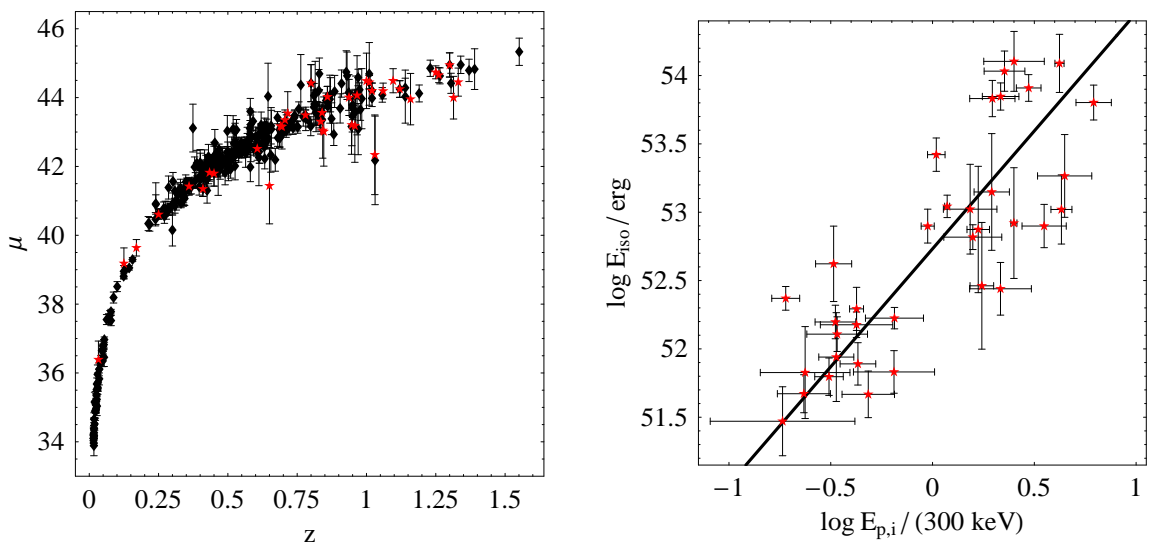}
 \caption{\label{fig1}
 Left panel: The Hubble diagram of 307 SNIa (black diamonds)
 and 38 low redshift GRBs (red stars) whose distance moduli
 are derived by using cubic interpolation. Right panel: 38 GRBs
 data (red stars) in the
 $\log E_{\rm p,i}\,/({\rm 300\,keV})-\log E_{\rm iso}\,/{\rm erg}$
 plane. The best-fit calibration line is also plotted. See text
 for details.}
 \end{figure}
 \end{center}


\vspace{-10mm} 


\section{Calibrating GRBs with Amati relation}\label{sec2}
In~\cite{r33}, the authors used 192 SNIa compiled by Davis
 {\it et al.}~\cite{r40} to interpolate the distance moduli of
 low redshift GRBs. Recently, the Supernova Cosmology Project (SCP)
 collaboration released their latest 307 SNIa
 dataset~\cite{r36,r37}. This so-called Union compilation is the
 currently largest SNIa dataset. Obviously, a larger SNIa dataset
 could bring a better interpolation. On the other hand,
 in~\cite{r33} the 69 GRBs compiled by Schaefer~\cite{r10} have
 been used. As in~\cite{r10}, the derived distance moduli of
 high redshift GRBs in~\cite{r33} are the weighted average of
 all available distance moduli from five GRB luminosity relations.
 For each GRB luminosity relation, only about a dozen GRBs
 at $z<1.4$ can be used to calibrate the corresponding
 relation~\cite{r33}. Instead, in the present work, we
 calibrate GRBs only with the Amati relation, so that we can
 use a larger GRBs dataset for single GRB luminosity relation. Here,
 we adopt the 70 GRBs compiled by Amati {\it et al.}~\cite{r41}.
 Notice that in~\cite{r42} there is another 76 GRBs compilation
 which heavily overlaps the one of~\cite{r41}. However, since the
 fluence data given in~\cite{r42} are not in the form of bolometric
 fluence, they are not convenient for our computing. As
 in~\cite{r33}, we choose $z=1.4$ to be the divide line to separate
 GRBs into two groups, since in the 307 SNIa dataset of~\cite{r36}
 there is only one SNIa (2003ak) whose $z>1.4$. In the 70 GRBs
 compiled by Amati {\it et al.}~\cite{r41}, there are 38 GRBs
 at $z<1.4$ and 32 GRBs at $z>1.4$. The maximum redshift of
 these 70 GRBs is $z=6.29$ for GRB~050904.

Several years ago, Amati {\it et al.} found the
 $E_{\rm p,i}-E_{\rm iso}$ correlation in GRBs as
 $E_{\rm p,i}=K\times E_{\rm iso}^m$ by using 12 GRBs with
 known redshifts~\cite{r12}, where
 $E_{\rm p,i}=E_{\rm p,obs}\times (1+z)$ is the cosmological
 rest-frame spectral peak energy; the isotropic-equivalent
 radiated energy is given by
 \be{eq1}
 E_{\rm iso}=4\pi d_L^2 S_{\rm bolo}(1+z)^{-1},
 \ee
 in which $S_{\rm bolo}$ is the bolometric fluence of gamma
 rays in the GRB at redshift $z$, and $d_L$ is the luminosity
 distance of the GRB. Later, Amati {\it et al.} have updated it
 in~\cite{r43} and~\cite{r41}. Up to now, some theoretical
 interpretations have been proposed for the Amati
 relation~\cite{r21}. It might be geometrical effects due to
 the jet viewing angle with respect to a ring-shaped emission
 region~\cite{r44}, or with respect to a multiple sub-jet model
 structure~\cite{r45}. An alternative explanation of the Amati
 relation is related to the dissipative mechanism responsible
 for the prompt emission~\cite{r46}. For convenience, similar
 to~\cite{r10}, we can rewrite the Amati relation as
 \be{eq2}
 \log\frac{E_{\rm iso}}{\rm erg}=\lambda+
 b\,\log\frac{E_{\rm p,i}}{\rm \,300\,keV\,}\,,
 \ee
 where $\log$ indicates the logarithm to base $10$, whereas
 $\lambda$ and $b$ are constants to be determined. In the
 literature, the Amati relation was calibrated with the
 $E_{\rm iso}$ computed by {\em assuming} a $\Lambda$CDM
 cosmology with particular model parameters. As mentioned
 above, this is cosmology dependent and the circularity
 problem follows. Here, we instead use the method proposed
 in~\cite{r33} to calibrate the Amati relation in a
 cosmology independent manner.

As the first step, we derive the distance moduli for the 38
 low redshift ($z<1.4$) GRBs of~\cite{r41} by using cubic
 interpolation from the 307 SNIa compiled in~\cite{r36}. We
 present the interpolated distance moduli $\mu$ of these 38
 GRBs in the left panel of Fig.~\ref{fig1}. The corresponding
 error bars are also plotted. As in~\cite{r33}, when the cubic
 interpolation is used, the error of the distance modulus $\mu$
 for the GRB at redshift $z$ can be calculated by
 $$\sigma_\mu=\left(\sum\limits_{i=1}^4 A_i^2
 \epsilon_{\mu,i}^2\right)^{1/2},~~~~~~~{\rm where}~~~~~
 A_i\equiv\left.\prod\limits_{j\not=i}\left(z_j-z\right)
 \right/\prod\limits_{j\not=i}\left(z_j-z_i\right),$$
 in which $j$ runs from 1 to 4 but $j\not=i$; on the other
 hand, $\epsilon_{\mu,i}$ are the errors of the nearby SNIa
 whose redshifts are $z_i$. Then, by using the well-known
 \be{eq3}
 \mu=5\log d_L+25,
 \ee
 one can convert distance modulus $\mu$ into luminosity
 distance $d_L$ (in units of Mpc). From Eq.~(\ref{eq1}) with
 the corresponding $S_{\rm bolo}$ given in Table~1
 of~\cite{r41}, we can derive the $E_{\rm iso}$ for these 38
 GRBs at $z<1.4$. We present them in the right panel of
 Fig.~\ref{fig1}, whereas the $E_{\rm p,i}$ for these 38 GRBs
 at $z<1.4$ are read from Table~1 of~\cite{r41}. Also, we
 present the errors for these 38 GRBs at $z<1.4$, by simply
 using the error propagation. From Fig.~\ref{fig1}, one can
 see that the intrinsic scatter is dominating over the
 measurement errors. Therefore, as in~\cite{r10,r33}, the
 bisector of the two ordinary least squares~\cite{r47} will
 be used. Following the procedure of the bisector of the two
 ordinary least squares described in~\cite{r47}, we find the
 best fit to be
 \be{eq4}
 b=1.725~~~~~~~{\rm and}~~~~~~~\lambda=52.7322,
 \ee
 with $1\sigma$ uncertainties
 \be{eq5}
 \sigma_b=0.010~~~~~~~{\rm and}~~~~~~~\sigma_\lambda=0.0065.
 \ee
 The best-fit calibration line Eq.~(\ref{eq2}) with $b$ and
 $\lambda$ in Eq.~(\ref{eq4}) is also plotted in the right
 panel of Fig.~\ref{fig1}.

\begin{center}
 \begin{figure}[htbp]
 \centering
 \includegraphics[width=0.55\textwidth]{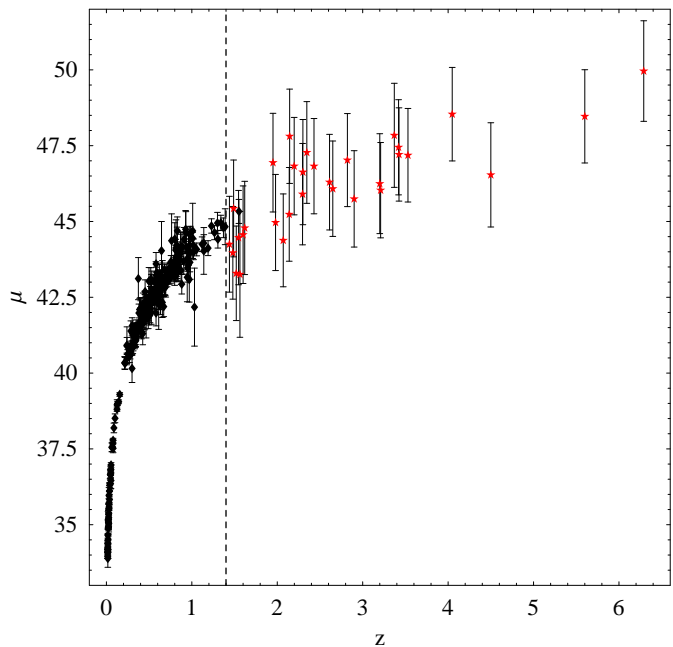}
 \caption{\label{fig2}
 The Hubble diagram of 307 SNIa (black diamonds)
 and 32 high redshift GRBs (red stars) whose distance moduli
 are derived by using the calibrated Amati relation. The dashed
 line indicates $z=1.4$.}
 \end{figure}
 \end{center}

Next, we extend the calibrated Amati relation to high redshift,
 namely $z>1.4$. Since the $E_{\rm p,i}$ for the 32 GRBs at $z>1.4$
 are given in Table~1 of~\cite{r41}, we can derive the $E_{\rm iso}$
 from the calibrated Amati relation Eq.~(\ref{eq2}) with $b$ and
 $\lambda$ in Eq.~(\ref{eq4}). Then, we derive the distance moduli
 $\mu$ for these 32 GRBs at $z>1.4$ using Eqs.~(\ref{eq1})
 and~(\ref{eq3}) while their $S_{\rm bolo}$ can be read from
 Table~1 of~\cite{r41}. On the other hand, the propagated
 uncertainties are given by~\cite{r10}
 \be{eq6}
 \sigma_\mu=\left[\left(\frac{5}{2}\sigma_{\log E_{\rm iso}}\right)^2
 +\left(\frac{5}{2\ln 10}\,\frac{\sigma_{S_{\rm bolo}}}{S_{\rm bolo}}
 \right)^2\right]^{1/2},
 \ee
 where
 \be{eq7}
 \sigma_{\log E_{\rm iso}}^2=\sigma_\lambda^2+\left(\sigma_b
 \log\frac{E_{\rm p,i}}{\rm \,300\,keV\,}\right)^2+\left(
 \frac{b}{\ln 10}\,\frac{\sigma_{E_{\rm p,i}}}{E_{\rm p,i}}\right)^2
 +\sigma_{E_{\rm iso,sys}}^2,
 \ee
 in which $\sigma_{E_{\rm iso,sys}}$ is the systematic error and it
 accounts the extra scatter of the luminosity relation. As
 in~\cite{r10}, by requiring the $\chi^2/dof$ of the 38 points at
 $z<1.4$ in the
 $\log E_{\rm p,i}\,/({\rm 300\,keV})-\log E_{\rm iso}\,/{\rm erg}$
 plane about the best-fit calibration line to be unity, we find that
 \be{eq8}
 \sigma_{E_{\rm iso,sys}}^2=0.184.
 \ee
 We present the derived distance moduli $\mu$ with $1\sigma$
 uncertainties for these 32 GRBs at $z>1.4$ in Table~\ref{tab1}
 and Fig.~\ref{fig2}. It is worth noting that they are obtained
 in a completely cosmology independent manner.


\vspace{-10mm} 


\section{Smoothly reconstructing the cosmic expansion history
 with~the~calibrated~GRBs}\label{sec3}
Following the well-known procedure which is frequently used in the
 analysis of large-scale structure~\cite{r48}, Shafieloo
 {\it et al.} proposed a new method in~\cite{r38} to smooth noisy
 data directly using a Gaussian smoothing function, rather than the
 top hat smoothing function. The iterative method calculates the
 $\ln d_L (z)$ as
 \be{eq9}
 \ln d_L(z)^s_n=\ln d_L(z)^s_{n-1}+N(z)\sum\limits_i\left[
 \ln d_L^{obs}(z_i)-\ln d_L(z_i)^s_{n-1}\right]\exp\left[
 -\frac{\ln^2\left(\frac{1+z}{1+z_i}\right)}{2\Delta^2}\right],
 \ee
 where $\Delta$ is a constant to be given priorly, and the
 normalization parameter $N(z)$ is given by
 \be{eq10}
 N(z)^{-1}=\sum\limits_i\exp\left[
 -\frac{\ln^2\left(\frac{1+z}{1+z_i}\right)}{2\Delta^2}\right].
 \ee
 In this section, following the method proposed in~\cite{r38,r39}, we
 smooth the 307 SNIa data and the 32 calibrated GRBs data at $z>1.4$
 to reconstruct the cosmic expansion history up to $z=6.29$.

As is well known, the luminosity distance
 \be{eq11}
 d_L=c\,(1+z)\int_0^z\frac{d\tilde{z}}{H(\tilde{z})},
 \ee
 where $H(z)$ is the Hubble parameter, and
 $c=2.9979\times 10^{10}~ \rm cm/s$ is the speed of light.
 For convenience, we introduce the dimensionless luminosity distance
 \be{eq12}
 D_L\equiv (1+z)\int_0^z\frac{d\tilde{z}}{E(\tilde{z})},
 \ee
 where $E\equiv H/H_0$ and $H_0=H(z=0)=100\,h~\rm km/s/Mpc$ is the
 Hubble constant. Thus, $d_L=cH_0^{-1}D_L$ and consequently
 \bea
 \mu&=&5\log d_L+25=5\left(\log\frac{cH_0^{-1}}{\rm Mpc}+
 \log D_L\right)+25\nonumber\\
 &=&5\log f+42.3841=\frac{5}{\ln 10}\ln f+42.3841,\label{eq13}
 \eea
 where $f\equiv D_L/h$ and hence $\ln D_L=\ln f+\ln h$. On the
 other hand, $\ln d_L=\ln f+\ln 2997.9$. Similar to~\cite{r39},
 we can rewrite Eq.~(\ref{eq9}) as
 \be{eq14}
 \ln f(z)^s_n=\ln f(z)^s_{n-1}+N(z)\sum\limits_i\left[
 \ln f^{obs}(z_i)-\ln f(z_i)^s_{n-1}\right]\exp\left[
 -\frac{\ln^2\left(\frac{1+z}{1+z_i}\right)}{2\Delta^2}\right],
 \ee
 where $\ln f^{obs}=(\mu^{obs}-42.3841)(\ln 10)/5$ from
 Eq.~(\ref{eq13}). When $n=1$,
 \bea
 \ln f(z)^s_1&=&\ln f(z)^s_0+N(z)\sum\limits_i\left[
 \ln f^{obs}(z_i)-\ln f(z_i)^s_0\right]\exp\left[
 -\frac{\ln^2\left(\frac{1+z}{1+z_i}\right)}{2\Delta^2}
 \right]\nonumber\\
 &=&\ln D_L(z)^s_0+N(z)\sum\limits_i\left[
 \ln f^{obs}(z_i)-\ln D_L(z_i)^s_0\right]\exp\left[
 -\frac{\ln^2\left(\frac{1+z}{1+z_i}\right)}{2\Delta^2}
 \right],\label{eq15}
 \eea
 where $D_L(z)^s_0$ is the dimensionless luminosity distance of
 the initial guess background model. The $\chi^2$ at any iteration
 is calculated as
 \be{eq16}
 \chi^2_n=\sum\limits_i\frac{\left[\mu(z_i)_n
 -\mu^{obs}(z_i)\right]^2}{\sigma^2_{\mu_{obs,i}}}.
 \ee
 The best fit is at the minimum $\chi^2_n$. Note that the summation
 is over the 307 SNIa and 32 calibrated GRBs at $z>1.4$. From
 Eq.~(\ref{eq11}) and $d_L=cH_0^{-1}D_L=fc/100$, we can find the
 Hubble parameter as
 \be{eq17}
 H(z)=\left\{\frac{d}{dz}\left[\frac{d_L(z)}{c\,(1+z)}\right]
 \right\}^{-1}
 =\left\{\frac{d}{dz}\left[\frac{f(z)}{100\,(1+z)}\right]
 \right\}^{-1}.
 \ee
 Then, the deceleration parameter $q(z)$ and the total
 equation-of-state parameter $w_{tot}\equiv p_{tot}/\rho_{tot}$
 can be given by
 \bea
 q(z)=(1+z)\frac{H^\prime(z)}{H(z)}-1,\label{eq18}\\
 w_{tot}(z)=-1+\frac{2}{3}(1+z)\frac{H^\prime(z)}{H(z)},\label{eq19}
 \eea
 where a prime denotes a derivative with respect to $z$. We
 intensively refer to the original papers~\cite{r38,r39} for
 technical details.

 \begin{table}[ptbh]
 \begin{center}
 \begin{tabular}{llccc}
 \hline\hline \ GRB & ~~~~~$z$ & ~~~~$S_{\rm bolo}~(10^{-5}~\rm erg\, cm^{-2}$)
 & ~~~~~$E_{\rm p,i}~(\rm keV)$ & \ \ $\mu$ \\ \hline
 \ 050318    & ~~~~ 1.44     & ~~~~ 0.42$\,\pm\,$0.03    & ~~~~ 115$\,\pm\,$25   & ~~~~ 44.25$\,\pm\,$1.58 \ \\
 \ 010222    & ~~~~ 1.48     & ~~~~ 14.6$\,\pm\,$1.5     & ~~~~ 766$\,\pm\,$30   & ~~~~ 43.97$\,\pm\,$1.53 \ \\
 \ 060418    & ~~~~ 1.489    & ~~~~ 2.3$\,\pm\,$0.5  & ~~~~ 572$\,\pm\,$143  & ~~~~ 45.43$\,\pm\,$1.60 \ \\
 \ 030328    & ~~~~ 1.52     & ~~~~ 6.4$\,\pm\,$0.6  & ~~~~ 328$\,\pm\,$55   & ~~~~ 43.29$\,\pm\,$1.56 \ \\
 \ 070125    & ~~~~ 1.547    & ~~~~ 13.3$\,\pm\,$1.3     & ~~~~ 934$\,\pm\,$148  & ~~~~ 44.47$\,\pm\,$1.56 \ \\
 \ 040912    & ~~~~ 1.563    & ~~~~ 0.21$\,\pm\,$0.06    & ~~~~ 44$\,\pm\,$33    & ~~~~ 43.26$\,\pm\,$2.08 \ \\
 \ 990123    & ~~~~ 1.6  & ~~~~ 35.8$\,\pm\,$5.8     & ~~~~ 1724$\,\pm\,$466     & ~~~~ 44.56$\,\pm\,$1.61 \ \\
 \ 990510    & ~~~~ 1.619    & ~~~~ 2.6$\,\pm\,$0.4  & ~~~~ 423$\,\pm\,$42   & ~~~~ 44.79$\,\pm\,$1.54 \ \\
 \ 080319C   & ~~~~ 1.95     & ~~~~ 1.5$\,\pm\,$0.3  & ~~~~ 906$\,\pm\,$272  & ~~~~ 46.94$\,\pm\,$1.63 \ \\
 \ 030226    & ~~~~ 1.98     & ~~~~ 1.3$\,\pm\,$0.1  & ~~~~ 289$\,\pm\,$66   & ~~~~ 44.97$\,\pm\,$1.59 \ \\
 \ 000926    & ~~~~ 2.07     & ~~~~ 2.6$\,\pm\,$0.6  & ~~~~ 310$\,\pm\,$20   & ~~~~ 44.38$\,\pm\,$1.53 \ \\
 \ 011211    & ~~~~ 2.14     & ~~~~ 0.5$\,\pm\,$0.06     & ~~~~ 186$\,\pm\,$24   & ~~~~ 45.24$\,\pm\,$1.55 \ \\
 \ 071020    & ~~~~ 2.145    & ~~~~ 0.87$\,\pm\,$0.4     & ~~~~ 1013$\,\pm\,$160     & ~~~~ 47.81$\,\pm\,$1.56 \ \\
 \ 050922C   & ~~~~ 2.198    & ~~~~ 0.47$\,\pm\,$0.16    & ~~~~ 415$\,\pm\,$111  & ~~~~ 46.83$\,\pm\,$1.61 \ \\
 \ 060124    & ~~~~ 2.296    & ~~~~ 3.4$\,\pm\,$0.5  & ~~~~ 784$\,\pm\,$285  & ~~~~ 45.90$\,\pm\,$1.67 \ \\
 \ 021004    & ~~~~ 2.3  & ~~~~ 0.27$\,\pm\,$0.04    & ~~~~ 266$\,\pm\,$117  & ~~~~ 46.63$\,\pm\,$1.73 \ \\
 \ 051109A   & ~~~~ 2.346    & ~~~~ 0.51$\,\pm\,$0.05    & ~~~~ 539$\,\pm\,$200  & ~~~~ 47.28$\,\pm\,$1.68 \ \\
 \ 060908    & ~~~~ 2.43     & ~~~~ 0.73$\,\pm\,$0.07    & ~~~~ 514$\,\pm\,$102  & ~~~~ 46.82$\,\pm\,$1.57 \ \\
 \ 050820    & ~~~~ 2.612    & ~~~~ 6.4$\,\pm\,$0.5  & ~~~~ 1325$\,\pm\,$277     & ~~~~ 46.30$\,\pm\,$1.58 \ \\
 \ 030429    & ~~~~ 2.65     & ~~~~ 0.14$\,\pm\,$0.02    & ~~~~ 128$\,\pm\,$26   & ~~~~ 46.08$\,\pm\,$1.57 \ \\
 \ 050603    & ~~~~ 2.821    & ~~~~ 3.5$\,\pm\,$0.2  & ~~~~ 1333$\,\pm\,$107     & ~~~~ 47.02$\,\pm\,$1.53 \ \\
 \ 050401    & ~~~~ 2.9  & ~~~~ 1.9$\,\pm\,$0.4  & ~~~~ 467$\,\pm\,$110  & ~~~~ 45.75$\,\pm\,$1.59 \ \\
 \ 020124    & ~~~~ 3.2  & ~~~~ 1.2$\,\pm\,$0.1  & ~~~~ 448$\,\pm\,$148  & ~~~~ 46.25$\,\pm\,$1.65 \ \\
 \ 060526    & ~~~~ 3.21     & ~~~~ 0.12$\,\pm\,$0.06    & ~~~~ 105$\,\pm\,$21   & ~~~~ 46.03$\,\pm\,$1.57 \ \\
 \ 030323    & ~~~~ 3.37     & ~~~~ 0.12$\,\pm\,$0.04    & ~~~~ 270$\,\pm\,$113  & ~~~~ 47.84$\,\pm\,$1.72 \ \\
 \ 971214    & ~~~~ 3.42     & ~~~~ 0.87$\,\pm\,$0.11    & ~~~~ 685$\,\pm\,$133  & ~~~~ 47.45$\,\pm\,$1.57 \ \\
 \ 060707    & ~~~~ 3.425    & ~~~~ 0.23$\,\pm\,$0.04    & ~~~~ 279$\,\pm\,$28   & ~~~~ 47.21$\,\pm\,$1.54 \ \\
 \ 060115    & ~~~~ 3.53     & ~~~~ 0.25$\,\pm\,$0.04    & ~~~~ 285$\,\pm\,$34   & ~~~~ 47.19$\,\pm\,$1.54 \ \\
 \ 060206    & ~~~~ 4.048    & ~~~~ 0.14$\,\pm\,$0.03    & ~~~~ 394$\,\pm\,$46   & ~~~~ 48.54$\,\pm\,$1.54 \ \\
 \ 000131    & ~~~~ 4.5  & ~~~~ 4.7$\,\pm\,$0.8  & ~~~~ 987$\,\pm\,$416  & ~~~~ 46.54$\,\pm\,$1.72 \ \\
 \ 060927    & ~~~~ 5.6  & ~~~~ 0.27$\,\pm\,$0.04    & ~~~~ 475$\,\pm\,$47   & ~~~~ 48.47$\,\pm\,$1.54 \ \\
 \ 050904    & ~~~~ 6.29     & ~~~~ 2.0$\,\pm\,$0.2    & ~~~~ 3178$\,\pm\,$1094    & ~~~~ 49.96$\,\pm\,$1.66 \ \\
 \hline\hline
 \end{tabular}
 \end{center}
 \caption{\label{tab1} The numerical data of 32 calibrated GRBs at
 $z>1.4$. The first 4 columns are read from Table~1 of~\cite{r41},
 whereas the last column is derived by using the calibrated Amati
 relation. See text for details.}
 \end{table}


 \begin{center}
 \begin{figure}[tbp]
 \centering
 \includegraphics[width=0.5\textwidth]{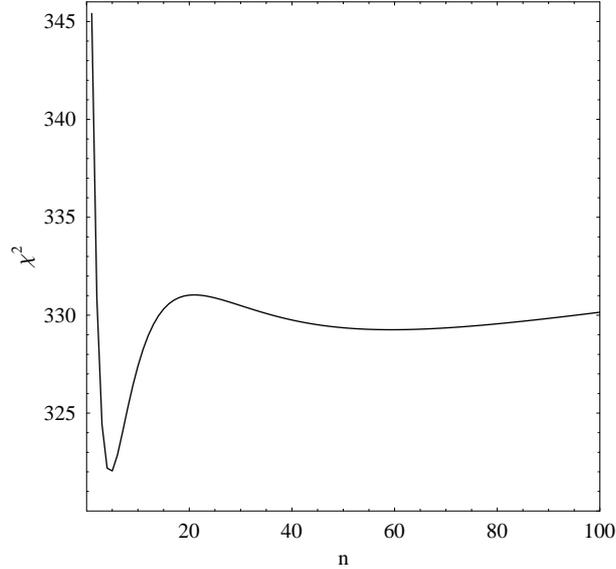}
 \caption{\label{fig3}
 Computed $\chi^2$ for the reconstructed results at each iteration
 using 307 SNIa and 32 calibrated high redshift GRBs at $z>1.4$.}
 \end{figure}
 \end{center}


\vspace{-11mm} 

\begin{center}
 \begin{figure}[htbp]
 \centering
 \includegraphics[width=0.93\textwidth]{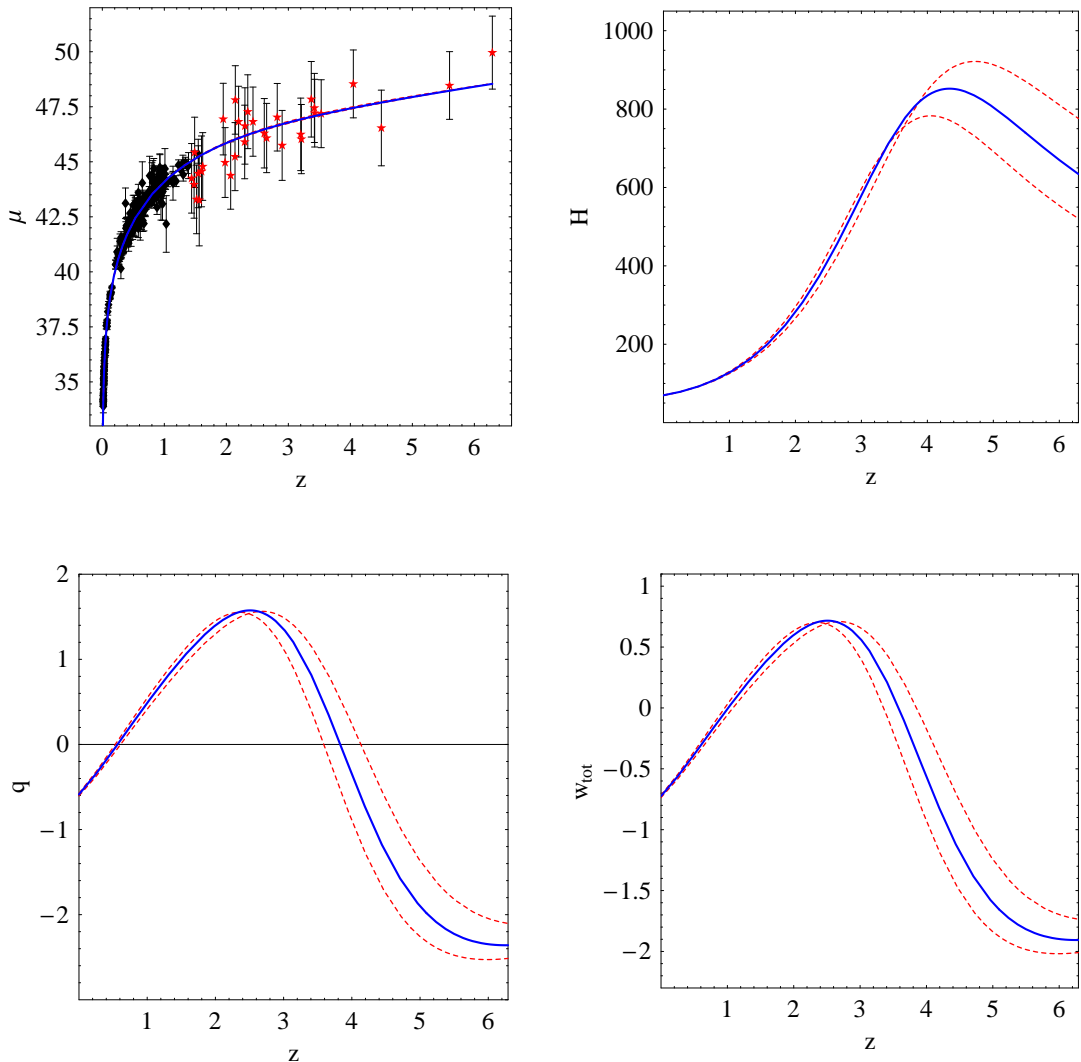}
 \caption{\label{fig4}
 The top-left panel is the three reconstructed $\mu(z)$ lines
 with the likelihood within $1\sigma$, whereas the best-fit
 result with $\chi^2_{min}$ is indicated by a blue solid line.
 In fact, these three reconstructed $\mu(z)$ lines
 cannot be significantly distinguished. We also plot the
 $\mu^{obs}$ of 307 SNIa (black diamonds) and 32 calibrated
 GRBs (red stars) in the $\mu(z)$ panel for comparison. From
 Eqs.~(\ref{eq17})---(\ref{eq19}), the $H(z)$, $q(z)$ and
 $w_{tot}(z)$ according to these three $\mu(z)$ lines are also
 plotted in the other panels. See text for details.}
 \end{figure}
 \end{center}

As shown in~\cite{r38}, the reconstructed results are {\em not}
 sensitive to the chosen value of $\Delta$ and the initial
 guess model. So, we choose $\Delta=0.55$ and use the flat
 $\Lambda$CDM model with $\Omega_{m0}=0.15$ as the initial
 guess model (notice that $\Omega_{m0}=0.15$ is the best fit
 of~\cite{r56} and \cite{r41}, which also fit $\Lambda$CDM
 model to GRBs data). For the flat $\Lambda$CDM model,
 $E(z)=\left[\Omega_{m0}(1+z)^3+(1-\Omega_{m0})\right]^{1/2}$.
 In Fig.~\ref{fig3}, we present the computed $\chi^2$ for the
 reconstructed results at each iteration using 307 SNIa and 32
 calibrated high redshift GRBs at $z>1.4$. It is easy to see
 that the $\chi^2$ goes to its minimum value very fast at just
 fifth iteration (similar to the case of the first reference
 in~\cite{r38}). The corresponding $\chi^2_{min}=322.045$. In
 the top-left panel of Fig.~\ref{fig4}, we present the three
 reconstructed $\mu(z)$ lines with the likelihood within
 $1\sigma$, whereas the best-fit result with $\chi^2_{min}$ is
 indicated by a blue solid line. In fact, these three reconstructed
 $\mu(z)$ lines cannot be significantly distinguished. From
 Eqs.~(\ref{eq17})---(\ref{eq19}), the $H(z)$, $q(z)$ and
 $w_{tot}(z)$ according to these three $\mu(z)$ lines are also
 plotted in the other panels of Fig.~\ref{fig4}. From
 Fig.~\ref{fig4}, we read that $H_0=68.817~\rm km/s/Mpc$,
 $q_0=-0.587$ and $w_{tot}(z=0)=-0.724$. Interestingly, while
 the results are familiar at relatively low redshifts which
 cover the redshift range of SNIa, the reconstructed $H(z)$,
 $q(z)$ and $w_{tot}(z)$ decrease at high redshifts which are
 in the redshift range of GRBs. The $w_{tot}$ crossed the
 phantom divide $-1$.


\vspace{-15mm} 


\section{Conclusion and discussions}\label{sec4}
In this work, we used the cosmology independent method proposed
 in~\cite{r33} to calibrate the GRBs and derived the distance
 moduli $\mu$ for 32 high redshift GRBs at $z>1.4$, whereas the
 numerical results are given in Table~\ref{tab1}. We have used
 the 307 SCP Union SNIa compilation~\cite{r36} and the 70 GRBs
 compiled in~\cite{r41}. Since these GRBs are calibrated in a
 completely cosmology independent manner, one can use the
 calibrated 32 GRBs at $z>1.4$ to constrain cosmological models
 without circularity problem completely. On the other hand,
 following the method proposed by Shafieloo {\it et al.}~\cite{r38},
 in this work we smoothly reconstructed the cosmic expansion history
 up to $z=6.29$ using 307 SNIa compiled in~\cite{r36} and 32
 calibrated GRBs at $z>1.4$. It is worth noting that this method is
 also model independent.

Here are some comments on the technical details~\cite{r74}.
 Firstly, SNIa as standard candles are affected by several
 potential problems, e.g., dust extinction, color evolution,
 reliability of the Phillips relation up to high $z$,
 uncertainty on the progenitors, possible existence of
 different sub-classes, etc. Thus, calibrating GRB with SNIa
 propagates these uncertainties and sources of systematics also
 into the calibrated spectrum-energy correlations. In addition,
 even when using a cosmology independent method, in this way
 GRBs are no more an ``independent'' cosmological probe. In
 other words, by calibrating with SNIa one can avoid the
 circularity problem intrinsic in the use of GRB
 spectrum-energy correlations and increase the accuracy in the
 determination of spectral parameters. However, with respect
 to using spectrum-energy correlations alone, one might
 introduce more systematics and introduce a ``circularity''
 with SNIa themselves. Secondly, in the present work, we used
 the bisector of the two ordinary least squares~\cite{r47} to
 derive the best-fit parameters of the calibrated
 $E_{\rm p,i}-E_{\rm iso}$ correlation. In fact, we closely
 followed the method used by Schaefer~\cite{r10}. In this
 method, one changed the value of $\sigma^2_{E_{\rm iso,sys}}$
 {\em by hand} to force the $\chi^2/dof$ of the 38 points at
 $z<1.4$ to be $1$. One might consider that this is not a
 robust method. As a good alternative, we refer to the
 maximum likelihood method used, e.g., by Amati
 {\it et al.}~\cite{r41}.

As mentioned above, the reconstructed $H(z)$, $q(z)$ and
 $w_{tot}(z)$ are interesting in some sense. To make these
 results robust, we have tried various values of $\Delta$ and
 various initial guess models. The resulted $H(z)$, $q(z)$
 and $w_{tot}(z)$ are still similar to the previous ones. The
 features in the reconstructed $H(z)$, $q(z)$ and $w_{tot}(z)$
 at high redshift $z>2$ remain. Some remarks are in order.
 Firstly, one might doubt the validity of using GRBs as
 standard candles. If GRBs are not real standard candles, the
 reconstructed results at $z>1.4$ are unreliable of course.
 In fact, the debate is not settled in the GRB community today.
 Some authors argued that GRBs cannot be used as standard candles,
 see~\cite{r49,r50,r51,r52,r53,r70,r71,r72,r73} for examples.
 On the other side, some authors advocated that GRBs
 can be used as standard candles to probe cosmology,
 see~\cite{r10,r11,r12,r13,r14,r15,r16,r17,r18,r19,r20,r21,r22,
 r23,r24,r25,r26,r27,r28,r29,r30,r31,r32,r33,r34,r35,r54,r55,
 r56,r57,r58,r59,r60,r61,r62,r63} for instance. Today, the
 situation is still unclear and many authors choose to wait and
 see. Secondly, if the GRBs can be used as standard candles,
 since the sample of available GRBs at high redshifts is still
 not large enough, some bias might exist in the reconstructed
 $H(z)$, $q(z)$ and $w_{tot}(z)$ and lead to the interesting
 features. So, before more and better high redshift GRBs are
 available, we cannot say anything conclusively. Finally, on
 the other side, some hints were found for oscillating $H(z)$,
 $q(z)$ and $w_{tot}(z)$ in the literature, see
 e.g.~\cite{r64,r65,r66,r67,r68} and references therein. So,
 the reconstructed results obtained here might be meaningful in
 some sense and deserve further investigations.

Obviously, due to the large scatter and the lack of a large
 amount of well observed GRBs, there is a long way to use GRBs
 extensively and reliably to probe cosmology. The Fermi
 Gamma-ray Space Telescope~\cite{r69} which was launched
 recently might change this situation. The GRB sample will be
 appreciably larger, and the new GRBs will be better studied
 in some sense. In the coming Fermi era, we hope that the GRB
 cosmology could have a bright future.


\section*{ACKNOWLEDGEMENTS}
We thank the anonymous referee for quite useful comments and
 suggestions, which helped us to improve this work. We are
 grateful to Prof.~Rong-Gen~Cai, Prof.~Xinmin~Zhang and
 Prof.~Zu-Hui~Fan for helpful discussions. We also thank
 Minzi~Feng, as well as Nan~Liang, Pu-Xun~Wu, Yuan~Liu,
 Wei-Ke~Xiao, Rong-Jia~Yang, Jian~Wang, Lin~Lin, Bin~Fan, and
 Bin~Shao, Yu~Tian, Zhao-Tan~Jiang, Feng~Wang, Jian~Zou,
 Zhi~Wang, Xiao-Ping~Jia, for kind help and discussions. This
 work was supported by the Excellent Young Scholars
 Research Fund of Beijing Institute of Technology.

\renewcommand{\baselinestretch}{1.1}


\end{document}